\documentclass{emulateapj}
\usepackage{amsmath}
\usepackage{graphicx}% Include figure files
\usepackage{subfigure} 
\usepackage{dcolumn}% Align table columns on decimal point
\usepackage{bm}% bold math
\usepackage{rotating}
\usepackage{hyperref}

\begin{document}

\title{Measurement of the Anisotropy of Cosmic Ray Arrival Directions with
IceCube}

\author{
IceCube Collaboration:
R.~Abbasi\altaffilmark{1},
Y.~Abdou\altaffilmark{2},
T.~Abu-Zayyad\altaffilmark{3},
J.~Adams\altaffilmark{4},
J.~A.~Aguilar\altaffilmark{1},
M.~Ahlers\altaffilmark{5},
K.~Andeen\altaffilmark{1},
J.~Auffenberg\altaffilmark{6},
X.~Bai\altaffilmark{7},
M.~Baker\altaffilmark{1}
% ...
}

\altaffiltext{1}{Dept.~of Physics, University of Wisconsin, Madison, WI 53706, 
%%%
USA; rasha.abbasi@icecube.wisc.edu, paolo.desiati@icecube.wisc.edu
%%%
}
\altaffiltext{2}{Dept.~of Subatomic and Radiation Physics, University of Gent, B-9000 Gent, Belgium}
\altaffiltext{3}{Dept.~of Physics, University of Wisconsin, River Falls, WI 54022, USA}
\altaffiltext{4}{Dept.~of Physics and Astronomy, University of Canterbury, Private Bag 4800, Christchurch, New Zealand}
\altaffiltext{5}{Dept.~of Physics, University of Oxford, 1 Keble Road, Oxford OX1 3NP, UK}
\altaffiltext{6}{Dept.~of Physics, University of Wuppertal, D-42119 Wuppertal, Germany}
\altaffiltext{7}{Bartol Research Institute and Department of Physics and Astronomy, University of Delaware, Newark, DE 19716, USA}
% ...

\begin{abstract}

%\linenumbers

We report the first observation of an anisotropy in the arrival direction 
of cosmic rays with energies in the multi TeV region in the Southern sky 
using data from the IceCube detector. Between June 2007 and March
2008, the partially-deployed IceCube detector was operated in a configuration
with 1320 digital optical sensors distributed over 22 strings at depths between
1450 and 2450 meters inside the Antarctic ice. IceCube is a neutrino detector,
but the data are dominated by a large background of cosmic ray muons.
Therefore, the background data are suitable for high-statistics studies of
cosmic rays in the Southern sky. The data include 4.3 billion muons produced by
downgoing cosmic ray interactions in the atmosphere; these events were
reconstructed with a median angular resolution of 3 degrees and a median
energy of $\sim20$~TeV. Their arrival direction distribution exhibits an anisotropy in
right ascension with a first harmonic amplitude of  $(6.4\pm0.2~$stat$. \pm 0.8~$syst$.)\times10^{-4}$.

\end{abstract}


\section{Introduction}
%\linenumbers

Long-term observations of cosmic ray muons by underground experiments have
demonstrated the presence of an anisotropy in the cosmic ray
intensity up to a few hundred GeV~\citep{nag}.  Recent underground and surface
array measurements of cosmic rays by the Tibet Array~\citep{ta},
Super-Kamiokande~\citep{sk} and Milagro~\citep{milagro} indicate that the
anisotropy persists into the TeV range.  

All of the TeV measurements were performed in the Northern hemisphere; so far, no
such measurement has been performed covering the entire Southern hemisphere at
median energies in the multi TeV region. With the deployment of the IceCube Neutrino Observatory at the South Pole, we have for the first time
measured the anisotropy at TeV energies in the Southern sky.
IceCube is primarily a neutrino detector but it is sensitive to the muons produced in
downward-going cosmic ray air showers. The observatory provides high-statistics measurements
of cosmic rays with median energy of $20$~TeV.

When completed in 2011, IceCube will comprise 5160 optical modules buried
1450 and 2450 meters below the surface of the polar ice sheet.  The modules are
physically connected to the surface by electronic umbilical lines, or
``strings,'' with 86 strings in total~\citep{dommb}. In this
paper, we use cosmic ray data recorded by the detector in its 22-strings
configuration (IC22) between June 2007 and March 2008 to produce the 
cosmic ray skymap of the Southern sky in the TeV range.

\section{Analysis}

During the IC22 physics run, cosmic ray events were observed at an average
trigger rate of about 550 Hz.  
The arrival direction is determined by a likelihood based reconstruction 
which is seeded with a fast online estimate of the arrival direction~\citep{like}. The 
likelihood based reconstruction is applied if twelve or 
more optical sensors on at least three strings were triggered by the event.
A total of $5.2\times10^{9}$ events satisfied the above conditions at an
average rate of $\sim240$ Hz.  Further selection criteria were applied to the
data to ensure good quality and stable runs. The final data set contains
$4.3\times10^{9}$ events with a total livetime of 226 days, a median angular
resolution of $3^{\circ}$, and a median energy per cosmic ray of $20$~TeV. The
energy scale was determined  with a standard cosmic ray simulation program,
CORSIKA~\citep{sim1}, using the SIBYLL hadronic interaction model (Version
2.1)~\citep{sim2} and the Poly-Gonato model for the composition and spectrum of
the primary cosmic rays~\citep{sim3}.

To evaluate physical anisotropies in the cosmic ray data set, it is necessary
to eliminate spurious effects which can mimic an anisotropy.  These
include local effects such as diurnal and seasonal variations of atmospheric
conditions, asymmetries in the detector geometry, and nonuniform detector
exposure to different regions of the sky.  Fortunately, the location of IceCube
at the South Pole is ideal to compensate for many effects that can impact
cosmic ray detectors in the middle latitudes. At the South Pole, the Southern
celestial sky is fully visible at any given time, providing complete and
uniform coverage. While the seasonal variation in the cosmic ray event rate is on the order of $\pm 10\%$~\citep{Tilav}, these variations are sufficiently slow to have no effect on the anisotropy. Rapid atmospheric changes which can affect the rate are rare and can be identified from the data.
 
\begin{figure}[h!]
  \begin{center}
  \includegraphics[width=0.5\textwidth]{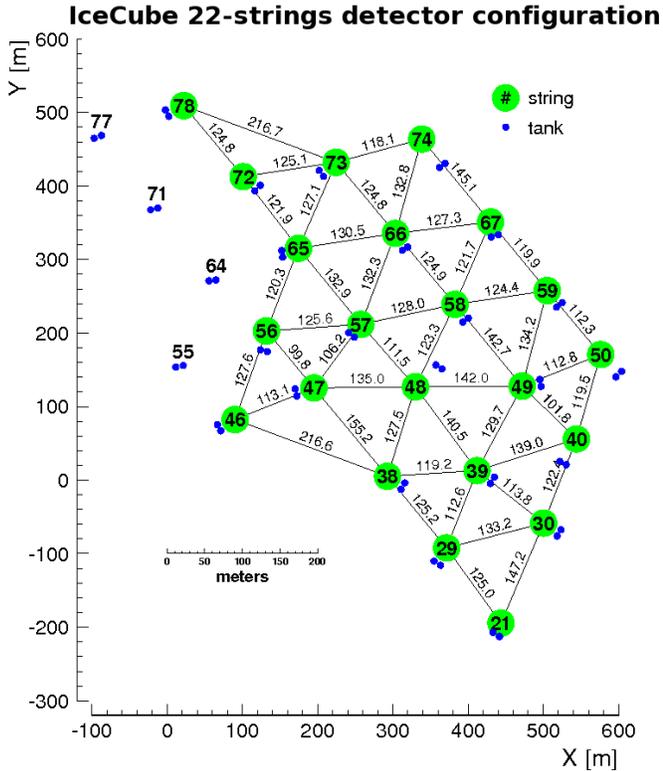}
  \caption{\label{configs} This plot shows the IceCube detector geometry in the 22 string configuration. The filled green circles are the positions of IceCube strings and the filled blue circles display the position of the IceTop tanks.}
  \end{center}
\end{figure}

The remaining effects which must be accounted for in this analysis are an
asymmetry in the IceCube detector response, and a non-uniformity in the time
coverage of the data.  The asymmetric response is due to the geometrical configuration of IceCube during the IC22 physics run (as shown in Figure~\ref{configs}); events
arriving along the long axis of the detector were preferentially selected by
the online filter and reconstruction due to the larger number of strings and
modules triggered.  In principle, the rotation of the Earth should average out
the local asymmetry in the arrival directions each day, but gaps in the
detector uptime and uneven run selection due to quality selection introduce
non-uniformities into the time coverage of the data.  These non-uniformities
preclude the complete averaging, and translate into an artificial arrival
direction asymmetry in equatorial coordinates.

% % %%%%%%%%%%%%%%%%%%%%%%%%%%%%%%%%%%%%%%%%%%%%%%%%%%%%%%%%%%%%%%%%%%
% %%%%%%%%%%%%%%%%%%%%%%%%%%%%%%%%%%%%%%%%%%%%%%%%%%%%%%%%%%%%%%%%%%

To correct for this detector-related asymmetry, each event from a given local
azimuth bin $i$ was weighted with the ratio $\bar{n}/n_i$, where $\bar{n}$
is the average number of events over the full range of local azimuths, and
$n_i$ is the number of events in local azimuth bin $i$. Since the local azimuth distribution
varies with zenith angle, the events were grouped into four zenith bands with
approximately equal numbers of events per band. The weighting is applied within
each band to remove the detector asymmetry.

\section{Results}

To investigate the arrival direction distribution of the cosmic rays, we studied
the relative intensity of the cosmic ray induced muon flux.  The arrival direction
distribution is dominated by the zenith angle dependence of the muon flux. The zenith
angle dependence is a result of varying overburden for the muons through 
the Antarctic ice. Therefore, the flux was normalized within 
declination belts of width $3^\circ$, which corresponds to the angular resolution 
of the data. This procedure provides the relative intensity of the event rate in each 
declination belt independently. 

\begin{figure}[h!]
  \begin{center}
  \includegraphics[width=0.5\textwidth]{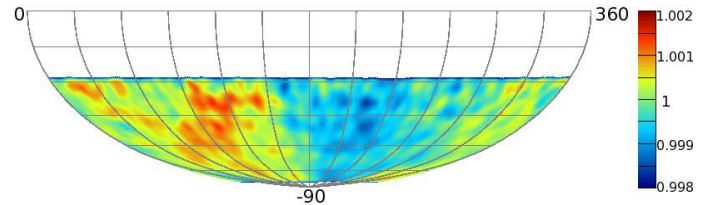}
  \caption{\label{sky}
  The relative intensity of the cosmic ray flux in equatorial coordinates.}
  \end{center}
\end{figure}

\begin{figure}[h!]
  \begin{center}
  \includegraphics[width=0.5\textwidth]{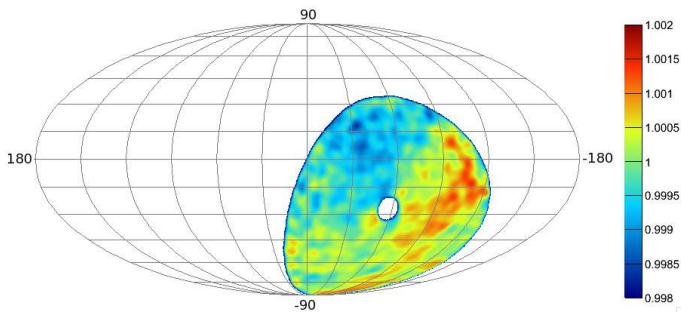}
  \caption{\label{gal_sky}
  The relative intensity of the cosmic ray flux in Galactic coordinates.}
  \end{center}
\end{figure}

 Figure~\ref{sky} shows the relative intensity of the event rate in equatorial coordinates. The color scale quantifies the number of reconstructed events with respect to the average number of events in each declination belt. Figure~\ref{gal_sky} shows the same data in Galactic
coordinates. Note that since the declination belts in the equatorial map
are treated independently, the map provides only information on the relative
modulation of the arrival direction of cosmic rays along the right ascension.  

\begin{figure}[h!]
  \begin{center}
  \includegraphics[width=0.5\textwidth]{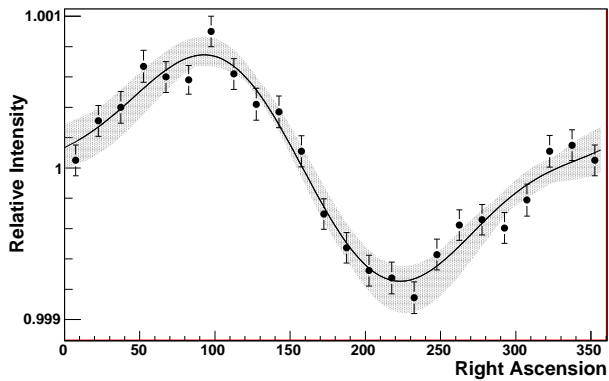}
  \caption{\label{1d}
  The one-dimensional projection in right ascension of the two-dimensional
  cosmic ray map in equatorial coordinates. The data are shown with 
  statistical uncertainties, and the black line corresponds to a fit to
  the data. The gray band indicates the estimated spread from the
  fit values of the stability tests (see text).}
  \end{center}
\end{figure}

Figure~\ref{sky} shows an anisotropy that appears to be a continuation
of a similar modulation of the cosmic ray flux observed in the Northern
hemisphere~\citep{ta,sk,milagro}. To quantify the scale of the anisotropy, 
we fitted the right ascension dependence of the data to a first- and second-order
harmonic function of the form
				  
\begin{equation}
  \sum_{i=1}^{n=2}A_{i}\cos(i(\alpha-\phi_{i})) + B,
  \label{eq1}
\end{equation}
where $(A_i,\phi_i)$ are the amplitude and phase of the anisotropy, $\alpha$ is
the right ascension, and $B$ is a constant.  Figure~\ref{1d} shows the
 anisotropy profile in right ascension obtained by accumulating the relative intensity distributions from the declination belts. The error bars are derived by propagating the statistical errors from each declination belt, and the gray band indicates the estimated spread from the
  fit values of the stability tests.  The solid line indicates the fit of equation~\eqref{eq1} to the data.  The first- and second-harmonic fit parameters to the one-dimensional projection in
Figure~\ref{1d} are $A_{1}=(6.4\pm0.2~$stat$. \pm 0.8~$syst$.)\times10^{-4}$,
$\phi_{1}=66.4^{\circ}\pm2.6^{\circ}~$stat$. \pm 3.8^\circ~$syst$.$, $A_{2}=(2.1\pm0.3~$stat$. \pm
0.5~$syst$.)\times10^{-4}$ and $\phi_{2}=-65.6^{\circ}\pm4.0^{\circ}~$stat$.\pm 7.5^\circ~$syst$.$, with
$\chi^{2}/\text{dof}=22/19$.

% % %%%%%%%%%%%%%%%%%%%%%%%%%%%%%%%%%%%%%%%%%%%%%%%%%%%%%%%%%%%%%%%%%%
% %%%%%%%%%%%%%%%%%%%%%%%%%%%%%%%%%%%%%%%%%%%%%%%%%%%%%%%%%%%%%%%%%%
% %%%%%%%%%%%%%%%%%%%%%%%%%%%%%%%%%%%%%%%%%%%%%%%%%%%%%%%%%%%%%%%%%%

% %%%%%%%%%%%%%%%%%%%%%%%%%%%%%%%%%%%%%%%%%%%%%%%%%%%%%%%%%%%%%%%%%%
% %%%%%%%%%%%%%%%%%%%%%%%%%%%%%%%%%%%%%%%%%%%%%%%%%%%%%%%%%%%%%%%%%%

To estimate the stability of the result and the corresponding systematic uncertainties we performed two types of tests: we checked whether the observation is stable against the choice of the particular event sample selection, and whether the modulation has spurious influences from other physical effects. The stability of the result was tested through a series of dedicated checks. The first stability test was done by dividing the data set in 
half  by sub-run number, where each sub-run contains approximately 20 minutes 
of observations. To avoid any systematic biases the division was tried in several 
ways: first, by separating data in even- and odd-numbered sub-runs, and second, 
by random selection of half of the sub-runs. The corresponding relative intensity 
distributions in right ascension for both tests were determined, and it was found 
that the variation induced by the data set selection are within the statistical fluctuations.
In addition, to check for daily variational effects, 
the data were divided in two sets: the first containing sub-runs with event rates above the median value for the corresponding 
day, and the second containing sub-runs with event rate below the median value. The corresponding 
relative intensity modulations were fit and found to be smaller than the statistical fluctuations. 
This means that a variation in the absolute event rate does not affect the modulation on arrival 
direction in right ascension.

Similarly, more stability tests were applied to check for effects due to seasonal variations and time gaps. To check for the seasonal effect data were divided into one set containing 
the winter months (June-October) and one set containing summer months (November-March). The relative
intensity variations in right ascension were fit and found to be consistent with the statistical fluctuations. 
To verify that the non-uniform time coverage due to missing sub-runs and other gaps in the data is 
correctly handled by the azimuthal re-weighting procedure (see Section 2), the relative intensity 
distribution from the full data set was compared with the one determined using only the days with 
minimal time gaps. The differences were found to be consistent with statistical fluctuations.

In each of the above stability tests, an independent fit was made to the relative intensity distribution as a 
function of right ascension using Equation~\ref{eq1}. The envelope from all the stability tests fit curves was constructed and it is shown as the gray band in Figure~\ref{1d}.

To verify whether the analysis procedure could induce a modulation in right ascension, the experimental event arrival directions were randomized to generate an isotropic distribution, and the same analysis was performed on this sample. The result was found to be consistent with isotropy.

\begin{figure}[h!]
  \begin{center}
  \includegraphics[width=0.5\textwidth]{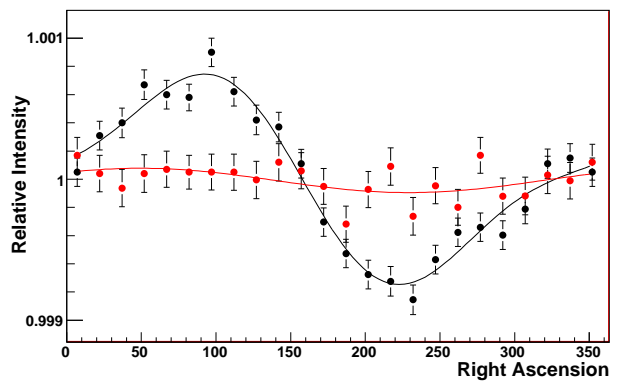}
  \caption{\label{anti}The one-dimensional projection in right ascension 
  for the data in sidereal time (in black) and in anti-sidereal time (in red). 
  The black line is the fit to the sidereal modulation using the first and 
  second harmonic fit of Equation~\ref{eq1}. The red line is the fit to the 
  anti-sidereal modulation using the first harmonic only.}
  \end{center}
\end{figure}

To check whether the observed anisotropy has some sidereal spurious effect derived from the interference between possible yearly-modulated daily variations, the same analysis was performed using the anti-sidereal time frame (a non-physical 
time defined by switching the sign of the transformation from universal to sidereal time)~\citep{antist}.
The real feature in the sidereal time is expected to be scrambled in the anti-sidereal time. Figure~\ref{anti}
shows the one dimensional projection in right ascension for the sidereal time in black
and for the anti-sidereal time in red. The amplitude of the first
harmonic fit to the one dimensional projection in the anti-sidereal
time was found to be $0.8\times10^{-4}$. This value is larger than the
spread found in the first harmonic amplitude from the stability tests,
therefore we use it as the systematic uncertainty in the first
harmonic amplitude. The uncertainty in the first harmonic
phase implied by the study in the anti-sidereal time frame
is within the systematic error determined from the stability tests.
The systematic uncertainties for the rest of the parameters
of the fit (quoted in Sec. 3) are derived from the
stability tests.

%%%%%%%%%%%%%%%%%%%%%%%%%%%%%%%%%%%%%%%%%%%%%%%%%%%%%%%%%%%%%%%%%%%
%%%%%%%%%%%%%%%%%%%%%%%%%%%%%%%%%%%%%%%%%%%%%%%%%%%%%%%%%%%%%%%%%%%

\section{Discussion}

Using a high-statistics sample of downgoing cosmic rays with a
median energy of $20$~TeV and a median angular resolution of $3^\circ$, we presented the first  map of the relative intensity of the flux of TeV
cosmic rays in the Southern sky.  The arrival direction distribution of the
cosmic rays is found to be anisotropic with a first harmonic amplitude and phase of $A_{1}=(6.4\pm0.2~$stat$. \pm
0.8~$syst$.)\times10^{-4}$ and $\phi_{1} = 66.4^\circ\pm2.6^\circ~$stat$. \pm 3.8^\circ~$syst$.$. The
observation appears to be a continuation of a previously-measured cosmic ray
anisotropy reported in the Northern hemisphere~\citep{ta,sk,milagro}. 

The origin of the anisotropy remains unclear. Compton-Getting~\citep{cg} 
suggested that the relative motion of the solar system around the Galactic 
center in the cosmic ray plasma should give rise to an excess in the
direction of motion of the solar system and a deficit in the opposite direction.  
In this model, an excess flux should appear with a maximum in right ascension between 
$290^{\circ}$ and $340^{\circ}$ and a minimum in right ascension  between $110^{\circ}$ and $160^{\circ}$~\citep{ta}. 
As shown in Fig 4, the excess can not be described in terms of the direction of motion 
of the solar system.  Therefore, we conclude that the Compton-Getting 
 effect could be (at most) one of several contributions to the cosmic ray anisotropy. 
This effect will be addressed in more detail in a future study including 
the energy dependence of the anisotropy.

It is tempting to try to interpret the cosmic ray excess as an artifact of the heliospheric magnetic field. However, the maximum gyro-radius of a 10 TeV cosmic ray proton in a 1 $\mu$G magnetic field is about 0.01 pc, i.e. much larger than the size of the heliosphere. As a consequence, the observed anisotropy is more likely to be connected to features of the local interstellar magnetic field at distances $<$ 1 pc. We are also investigating the possibility that the cosmic ray excess is associated with structures in the Galactic magnetic field at larger distance scales, or with diffusive particle flows from a nearby Galactic source such as Vela.

The still growing IceCube observatory will be completed in 2011 with a total of 86 strings and a  volume of 1 km$^3$. The estimated rate of cosmic ray-induced muons will be greater than  30 billion events per year. Such high statistical power, together with an estimated energy resolution of about 0.3 in log(E), will allow us to determine, in one year, the variation of cosmic ray anisotropy in several energy ranges up to a few hundred TeV. The energy dependence study will provide fundamental  hint at the nature of the source or sources of the cosmic rays, as well as their propagation through the Galactic magnetic field.

\section{Acknowledgements}
We acknowledge the support from the following agencies: U.S. National Science Foundation-Office of Polar Program, U.S. National Science Foundation-Physics Division, University of Wisconsin Alumni Research Foundation, U.S. Department of Energy, and National Energy Research Scientific Computing Center, the Louisiana Optical Network Initiative (LONI) grid computing resources; Swedish Research Council, Swedish Polar Research Secretariat, Swedish National Infrastructure for Computing (SNIC), and Knut and Alice Wallenberg Foundation, Sweden; German Ministry for Education and Research (BMBF), Deutsche Forschungsgemeinschaft (DFG), Research Department of Plasmas with Complex Interactions (Bochum), Germany; Fund for Scientific Research (FNRS-FWO), FWO Odysseus programme, Flanders Institute to encourage scientific and technological research in industry (IWT), Belgian Federal Science Policy Office (Belspo); Marsden Fund, New Zealand; Japan Society for Promotion of Science (JSPS); the Swiss National Science Foundation (SNSF), Switzerland; A. Kappes and A. Gro{\ss} acknowledge support by the EU Marie Curie OIF Program; J. P. Rodrigues acknowledge support by the Capes Foundation, Ministry of Education of Brazil. 

%\bibliographystyle{apj}
%\bibliography{ic22LSA_letter}

\begin{thebibliography}{12}
\expandafter\ifx\csname natexlab\endcsname\relax\def\natexlab#1{#1}\fi

\bibitem[{Abbasi {et~al.}(2009)}]{dommb}
Abbasi, R., {et~al.} 2009, Nucl. Instrum. Meth., A, 601, 294

\bibitem[{Abdo {et~al.}(2009)}]{milagro}
Abdo, A., {et~al.} 2009, Astrophys. J., 698, 2121

\bibitem[{Ahrens {et~al.}(2004)}]{like}
Ahrens, J., {et~al.} 2004, Nucl. Instrum. Meth., A, 524, 169

\bibitem[{Amenomori {et~al.}(2006)}]{ta}
Amenomori, M., {et~al.} 2006, Science, 314, 439

\bibitem[{Compton \& Getting(1935)}]{cg}
Compton, A.~H., \& Getting, I.~A. 1935, Physical Review, 47, 817

\bibitem[{CORSIKA(2009)}]{sim1}
CORSIKA. 2009, http://www-ik.fzk.de/corsika/

\bibitem[{Engel(1999)}]{sim2}
Engel, R. 1999, in International Cosmic Ray Conference, Vol.~1, 415

\bibitem[{Farley {et~al.}(1954)}]{antist}
Farley, F., {et~al.} 1954, in Physical Society. A., Vol.~67, 996

\bibitem[{Guillian {et~al.}(2007)}]{sk}
Guillian, G., {et~al.} 2007, Physical Review D, 75, 062003

\bibitem[{{H{\"o}randel}(2003)}]{sim3}
{H{\"o}randel}, J.~R. 2003, Astroparticle Physics, 19, 193

\bibitem[{Nagashima {et~al.}(1998)}]{nag}
Nagashima, K., {et~al.} 1998, Journal of Geophysical Research, 103, 17429

\bibitem[{Tilav {et~al.}(2010)}]{Tilav}
Tilav, S., {et~al.} 2010, ArXiv:astro-ph/1001.0776

\end{thebibliography}

\end{document}